\documentclass[prl,twocolumn,epsf,psfig]{revtex4}
\usepackage{graphicx}
\DeclareGraphicsExtensions{.pdf,.png,.gif,.jpg}
\usepackage{float}
\usepackage{subfig}
\usepackage{epsfig}
\usepackage{wrapfig}
\usepackage{bbold}

\begin{document}

\title{Asymptotic behavior of correlation functions of one-dimensional polar-molecules on optical lattices}

\author{Theja N. De Silva}
\affiliation{Department of Chemistry and Physics,
Augusta University, Augusta, Georgia 30912, USA;\\
Kavli Institute for Theoretical Physics, University of California, Santa Barbara, California 93106, USA.}

\begin{abstract}
We combine a slave-spin approach with a mean-field theory to develop an approximate theoretical scheme to study the density, spin, and, pairing correlation functions of fermionic polar molecules. We model the polar molecules subjected to a one-dimensional periodic optical lattice potential using a generalized $t-J$ model, where the long-range part of the interaction is included through the exchange interaction parameter. For this model, we derive a set of self-consistent equations for the correlation functions, and evaluate them numerically for the long-distance behaviour. We find that the pairing correlations are related to spin correlations through the density and the slave-spin correlations. Further, our calculations indicates that the long-range character of the interaction can be probed through these correlation functions. In the absence of exact solutions for the one-dimensional $t-J$ model, our approximate theoretical treatment can be treated as a useful tool to study one dimensional long-range correlated fermions. \end{abstract}

\maketitle

\section{I. Introduction}

Recent developments in laser technology and current extraordinary experimental advances in trapping and manipulating of ultra-cold atomic gases provide a wonderful path for the quantum simulation of many-body physics. Due to the unprecedented control over various experimental parameters, such as spatial dimensionality, lattice structures and geometries, interaction parameters, and atomic species, ultra-cold atomic systems are considered as promising playgrounds for studying fundamental condensed matter phenomena~\cite{REV1, REV2, REV3}. After the first generation of experimental systems of ultra-cold bosons and fermions with short-range interactions on optical lattices~\cite{REV4, REV5, REV6, REV7, REV8, REV9}, the focus has been shifted towards the polar-molecules as their long-range dipolar interactions give rise to more exotic many-body phenomena~\cite{REV9, REV10, REV11}. Unlike condensed matter counter part with Coulomb interactions, the dipolar-type long-range interactions are not affected by screening. As a result, the dynamics of polar molecules are described by tunable extended Hubbard-type Hamiltonians~\cite{model1, model2, model3}. The tunability of all Hamiltonian parameters can be achieved by manipulating the anisotropic dipole-dipole interactions between the homonuclear and heteronuclear molecules via external dc and ac fields in the microwave regime~\cite{mwf1, mwf2, mwf3, mwf4, mwf5, mwf6}. Some of the predicted exciting novel many-body phenomena due to the effect of long-range character and the anisotropy of the dipolar interactions~\cite{mbp1, mbp2, mbp3, mbp4, mbp5, mbp6, mbp7, mbp8, mbp8A}, have already been investigated in experiments~\cite{embp1, embp2, embp3}.

In this work, we study density, spin, and pairing correlation functions of a one-dimensional chain of polar molecules in a lattice. The dynamics of the polar molecules on the lattice can be represented by a generalized $t-J$ model with long range interactions between molecules in the form $1/r^3$, where $r$ is the spatial distance between molecules. The usual nearest neighbor $t-J$ model without this long range interaction part is the strong coupling limit of the well-known Hubbard model away from half filling. The strong coupling limit of the Hubbard model is defined as the limit $U \gg t$, where $U$ is the on-site interaction and $t$ is the hopping or tunneling amplitude. There is no exact solutions even for this nearest neighbor only one dimensional $t-J$ model, except for special cases~\cite{extj1, extj2, extj3}. The strong coupling limit of the Hubbard model $U \gg t$ is equivalent to the $t-J$ model when  $J \ll t$, where $J$ is the exchange spin coupling of the $t-J$ model. The exact Bethe-ansatz solutions for the nearest neighbor only one dimensional model is available only for $J \ll t$ limit~\cite{extj1} and $J = 2 t$ limit~\cite{extj2,extj3}. Even for these limits, the Bethe-ansatz wave functions provide limitations for the calculation of physical quantities due to the complexity of Bethe-ansatz solutions. Therefore, the study of the system of long-range interacting one dimensional lattice polar molecules requires heavily numerical or advanced novel theoretical techniques. It is the purpose of this paper to develop such a technique to tackle polar interacting molecules on optical lattices. We use a constraint free slave-spin approach to represent the quasi-particle operators and transform the interacting Hamiltonian into a combined slave spin and spinless fermion model. In this slave-spin representation, the original quasi-particle operators are decomposed into a bosonic pseudo-spin field and a fermionic field. However unlike other slave-boson representations, the charge degrees of freedom and the spin degrees of freedom are combinations of these two fields. We then evaluate the density, spin, and pairing correlation functions after decoupling the slave-spin and the spinless fermion sectors by using a mean-field theory. These correlation functions need to be derived self-consistently due to the coupling of pseudo-spin field and fermion field through their mean-field values. Our decoupled spinless fermion part of the Hamiltonian has the from of topological Hamiltonians presented in Refs.~\cite{CP1, CP2, CP3, CP4, CP5}. However, the effective interaction parameters in our transformed fermion part needs to be calculated self consistently in combination with slave-spin sector. We find that the zero temperature correlation functions show characteristic oscillatory decay, where oscillatory part originates from the pseudo-spin and spinless fermion correlations.

The paper is organized as follows. In section II, we introduce an effective model for the polar molecular system. The model is a generalization of the well known $t-J$ model, modified to include the long-range dipolar interactions. In section III, we introduce a slave-spin approach and convert our model Hamiltonain into a coupled spin-particle Hamiltonian. In section IV, we use a mean-field decoupling scheme to decouple the pseudo spin and spinless particle sectors of the Hamiltonian. In section V and VI, we solve the spin and spinless fermions parts independently in momentum space. In section VII, we combined the solutions of two sectors and derive self-consistent equations for the unknown mean field parameters. We dedicate section VIII to discuss the correlation functions and to provide our results. Finally, we present our conclusions with a short discussion in section IX.

\section{II. The model}

Our model describing the polar molecules in optical lattices is given by~\cite{model2},

\begin{eqnarray}
H = -t\sum_{\langle ij \rangle s}(c^\dagger_{i,s}c_{j,s}) \\ \nonumber
+\sum_{i < j} \frac{1}{|i-j|^3} \biggr[J_\perp (S^+_iS^{-}_j + S^{-}_iS^+_j) +J_z S^z_iS^z_j \biggr].
\end{eqnarray}

\noindent As usual, $c^\dagger_{i,s} (c_{i,s})$ are fermionic creation (annihilation) operator for a fermionic polar molecule with spin $s = \uparrow, \downarrow$ on lattice site $i$. The components of spin operators are defined as $S^+_i = c^\dagger_{i,\uparrow} c_{i,\downarrow} $, $S^{-}_{i} = c^\dagger_{i,\downarrow} c_{i,\uparrow}$, and $S^z_i = (c^\dagger_{i,\uparrow}c_{i, \uparrow} - c^\dagger_{i,\downarrow}c_{i, \downarrow})/2$. When deriving this model, it has been assumed that double occupations at a single site are not allowed due to the strong on-site interactions, thus the fermionic Hilbert space is projected onto the space with no doublons. The model is a straight forward generalization of the well-known $t-J$ model proposed for condensed matter systems~\cite{tj1, tj2, tj3, tj4}. All model parameters in our Hamiltonian and the average density per site can be controlled independently. Here we restrict ourselves to a simple experimental realization of the model by setting $J_z = 2 J_\perp \equiv J$ and assume hopping is restricted only to the nearest neighbors.

\section{III. Slave spin representation of the model}

Slave-particle approaches are very common in studying strongly correlated systems due to their simplicity in applying computational techniques and capability of accounting particle correlations beyond standard mean-field theories. While most variational approaches are valid only at zero temperature, the mean-field theories are unable to capture quantum fluctuations. However, slave particle theories are valid at both zero and finite temperatures and capable of capturing quantum fluctuations~\cite{spt1, spt2, spt3, spt4, spt4TD}. Further, it has been shown that slave-particle approaches are equivalent to a statistically-consistent Gutzwiller approximation~\cite{spg}. In slave particle formalisms, the original local Fock space of the problem is usually mapped onto a larger local Fock space that contains more states due to the introduction of auxiliary particles. In general these nonphysical states are removed in enlarge Hilbert space by imposing constraints. These constraints introduce additional self-consistent equations for the calculation. However in this work, we apply constraint-free, invertible canonical transformation proposed in Ref.~\cite{BK} to the correlated polar molecules on optical lattices. The transformation is  more effective than other slave-particle transformations as the basis states of the Hilbert space of a molecule on a single site has one-to-one mapping. This one-to-one mapping excludes the additional constraint equations in this slave-spin scheme. In this representation, the quasi-particle is described as a composite of \emph{spin-ness} and \emph{Fermi-ness}. While the spin-ness is described by a Pauli operator, the Fermi-ness is described by a spinless fermion~\cite{BK}. The physical spin and the physical particle number are related to both Pauli operators and spinless fermion number.

In this slave-spin approach, the particle operator is decoupled into a spinless fermion and a Pauli operator that carry the charge and spin degrees
of freedom, respectively. First, the quasi-particle operator $c^\dagger_{i,\uparrow/\downarrow}$ that creates an atom with spin $\uparrow/\downarrow$ at site $i$ is expressed as

\begin{eqnarray}
c^\dagger_{i,\uparrow} = (a^\dagger_i + a_i)\sigma^{+}_i/2,
\end{eqnarray}

\noindent and
\begin{eqnarray}
c^\dagger_{i,\downarrow} = [a^\dagger_i (1 - \sigma^z_i) - a_i(1 + \sigma^z_i)]/2.
\end{eqnarray}

\noindent
Notice that a typo of missing factor 1/2 in Ref.~\cite{BK} is corrected in Eq. (2). Under this transformation, the original number operators transform as $\hat{n}_{i,\uparrow} = (1 + \sigma^z_i)/2$ and $\hat{n}_{i,\downarrow} = 1/2 + (1/2 - \hat{n}) \sigma^z_i$, where $\hat{n} = a^\dagger_i a_i$ is the spinless fermion number operator. The physical spin $\vec{S}_i$ takes the form $\vec{S}_i = \hat{n}_i \vec{\sigma}_i/2$. Furthermore, the Hubbard-type interaction $(\hat{n}_{i,\uparrow}-1/2) (\hat{n}_{i,\downarrow} -1/2) = (1/2 - \hat{n})/2$. Therefore, the new Pauli operator $\vec{\sigma}$ represents the spin of the particles in the presence of spinless fermions and the strongly correlated nature of the original quasi-particles are captured by that of the spinless fermions. Further, the physical pairing operator whose components are given by $P^z_i = (\hat{n}_{i,\uparrow} + \hat{n}_{i,\downarrow} - 1)/2$ and $P^{\dagger}_i = c^\dagger_{i,\uparrow} c^\dagger_{i,\downarrow}$ has the form in new representation $\vec{P}_i = (\hat{n}_i-1)\vec{\sigma}_i/2$. Notice that in general the slave-particle transformation is constraint free. However, additional constraints need to be included in our model due to fact that no double occupancies of the spinfull fermions allowed on lattice sites. This condition simply indicates that states with $\sigma^z_i = +1$ and $n = 0$ are not allowed in the Hilbert space. In terms of new variables, our Hamiltonian becomes,

\begin{eqnarray}
H = -\frac{t}{2}\sum_{\langle ij \rangle}(a^\dagger_{i}a_{j} + h.c) (1 + \vec{\sigma}_i \cdot \vec{\sigma}_j) \\ \nonumber
-\frac{t}{2}\sum_{\langle ij \rangle}(a^\dagger_{i}a^\dagger_{j} + h.c) (1 - \vec{\sigma}_i \cdot \vec{\sigma}_j) \\ \nonumber
+\frac{J}{4} \sum_{i < j} \frac{1}{|i-j|^3} \hat{n}_i \hat{n}_j \vec{\sigma}_i \cdot \vec{\sigma}_j,
\end{eqnarray}

\noindent where the term "\emph{h.c}" stands for Hermitian conjugate. The last term is simply the $\vec{S}_i \cdot \vec{S}_j \rightarrow \hat{n}_i \hat{n}_j  \vec{\sigma}_i \cdot \vec{\sigma}_j/4$ and the first two terms originates from the transformed hopping part of the Hamiltonian.

\section{IV. Decoupling spin and fermions}

Due to the no double occupancy constraint of spinful fermions, the spinless fermions and slave-spins are strongly correlated. However, we believe that the essential physics can be captured by a simple mean-field decoupling scheme. We decouple the transformed Hamiltonian by using a mean-field description. By introducing four local mean-field parameters, $\langle a^\dagger_i a_j +h.c \rangle_a = \chi_{ij}$, $\langle a^\dagger_i a^\dagger_j +h.c \rangle_a = \Delta_{ij}$, $\langle \hat{n}_i \hat{n}_j \rangle_a = D_{ij}$, and $\langle \sigma_i \cdot \sigma_j\rangle_s = m_{ij}$, our Hamiltonian $H$ becomes the sum of independent spin and fermion parts: $H \rightarrow H_s + H_a$ . This will leads to the $H_s$ part being an interacting pure spin model and the $H_a$ part being an interacting spinless fermion part.  The subscript $a$ or $s$ means that the quantum and thermal expectation values must be taken with respect to the spin and fermion sectors, respectively. After performing the decoupling scheme, the fermion and spin parts of the Hamiltonian become,

\begin{eqnarray}
H_a = -\frac{t}{2}\sum_{\langle ij \rangle}(1 + m_{ij} )(a^\dagger_{i}a_{j} + h.c) \\ \nonumber
-\frac{t}{2}\sum_{\langle ij \rangle}(1-m_{ij})(a^\dagger_{i}a^\dagger_{j} + h.c)  \\ \nonumber
+\frac{J}{4} \sum_{i < j} \frac{m_{ij}}{|i-j|^3} \hat{n}_i \hat{n}_j,
\end{eqnarray}

\begin{eqnarray}
H_s = -\frac{t}{2}\sum_{\langle ij \rangle}(\chi_{ij} - \Delta_{ij})\vec{\sigma}_i \cdot \vec{\sigma}_j \\ \nonumber
+\frac{J}{4} \sum_{i < j} \frac{D_{ij}}{|i-j|^3}  \vec{\sigma}_i \cdot \vec{\sigma}_j \\ \nonumber
-\frac{t}{2}\sum_{\langle ij \rangle} (\chi_{ij} + \Delta_{ij}).
\end{eqnarray}

\section{V. Solution of the spin part}

By defining $|i-j| = l d_l$ with lattice constant $d_l$ and integer $l$, and rearranging the dummy variables in the sum, the spin part of the Hamiltonian can be casted as,

\begin{eqnarray}
H_s = -\sum_{l}^r \sum^{L-1}_{j = 1} J^e_l \vec{\sigma}_j \cdot \vec{\sigma}_{j+l}.
\end{eqnarray}

\noindent Here, we define $J^e_l = K_l \delta_{l1} +J_l$, where $K_l = t(\chi_l - \Delta_l)/2$ and $J_l = J D_l/(4 l^3 d_l^3)$, where we used a compact notation for local mean-field parameters, $\chi_{ij} = \chi_l$ for $|i-j| = l d_l$ etc. The parameter $L$ is the number of lattice sites, $r$ in general represents the range of interaction and $\delta_{l1}$ is the usual discrete Kronecker delta function. Notice that the decoupled pseudo spin part of the Hamiltonian is still an interacting long-range spin Hamiltonian. The classical ground state of this one-dimensional effective pseudo spin Hamiltonian on a Bravais lattice is a single-$\vec{Q}$ spiral state. Therefore, first we make a coordinate transformation by rotating the local axis by an angle $\theta_i = \vec{Q }\cdot \vec{r}_i$, such that a new axis-$\xi$ coincides with the classical solution of the pseudo-spin orientation,

\begin{eqnarray}
\sigma_i^x = \sigma_i^\eta \nonumber \\
\sigma_i^y = \sigma_i^\mu \cos \theta_i + \sigma_i^\xi \sin \theta_i \nonumber \\
\sigma_i^z = -\sigma_i^\mu \sin \theta_i + \sigma_i^\xi \cos \theta_i.
\end{eqnarray}

\noindent The resulting Hamiltonian then becomes,

\begin{eqnarray}
H_s = -\sum_{l}^r \sum^{L-1}_{j = 1} J^e_l \{\sigma_j^\eta \sigma_{j+l}^\eta + \sin (l Q d_l) [\sigma_j^\mu \sigma_{j+l}^\xi - \sigma_j^\xi \sigma_{j+l}^\mu] \nonumber \\
+ \cos (l Q d_l) [\sigma_j^\mu \sigma_{j+l}^\mu + \sigma_j^\xi \sigma_{j+l}^\xi]\}.
\end{eqnarray}

\noindent Representing pseudo spin operators by bosonic Holstein-Primakoff local operators, $\sigma_i^\eta = B^\dagger_i + B_i$, $\sigma_i^\mu = B^\dagger_i -B_i$, and $\sigma_i^\xi = 1 -  2 B_i^\dagger B_i$, and restricting ourselves to the quadratic order, the effective pseudo spin Hamiltonian in Fourier space can be written as,

\begin{eqnarray}
H_s = -\frac{1}{2}\sum_k \{2 J^e_l(k) [B_k B_{-k} + B^\dagger_k B^\dagger_{-k}] - 8 J^e_l(Q) B^\dagger_k B_k \nonumber \\
+ [2 J^e_l(k) + J^e_l(k+Q) + J^e_l(k-Q)][B_kB^\dagger_k + B^\dagger_k B_k] \},
\end{eqnarray}

\noindent where $ J^e(k) = \sum_l J^e_l e^{i lkd_l}$ is the Fourier transform of the effective coupling constant. The Hamiltonian can be brought to a diagonal form by usual Bogoliubov transformation, $A^\dagger_k = \cos \alpha_k B^\dagger_k + \sin \alpha_k B_k$,

\begin{eqnarray}
H_s = \sum_k \omega_k \biggr(A^\dagger_k A_k + \frac{1}{2}\biggr),
\end{eqnarray}

\noindent where the effective spin wave dispersion $\omega_k = \sqrt{M^2_{11}(k) - M^2_{21}(k)}$. Here the matrix elements of the Hamiltonian matrix $M_{11}(k)$ and $M_{12}(k)$ are given by, $M_{11}(k) = 2 J^e(Q) - J^e(k) -(1/2)[J^e(k+Q) + J^e(k-Q)]$ and $M_{12}(k) = (1/2)[J^e(k+Q) + J^e(k-Q)]-J^e(k)$.

\noindent The diagonal form of the Hamiltonian allows us to calculate the local mean field parameter, $m_{ij} = \langle \vec{\sigma}_i \cdot \vec{\sigma}_j\rangle_s$ through the Bogoliubov quasi particles. Assuming $j = i + l$ and we find $m_{ij} \equiv m_l = \sum_k e^{ikld_l} m_k$ through the Fourier transform $m_k = X^{-} (k) + [X^{+}(k-Q) + X^{+}(k+Q)]/2$, where

\begin{eqnarray}
X^{\pm}(k) = \frac{M_{11}(k) \pm M_{12}(k)}{\omega_k} (1 + n_k),
\end{eqnarray}

\noindent and $n_k = (e^{\hbar \omega_k/k_BT} -1)^{-1}$ is the Bogoliubov quasi particles boson occupation number, with the Boltzmann constant $k_B$, the planck constant $h = 2 \pi \hbar$, and temperature $T$.

\section{VI. Solution of the fermion part}

First we tackle the spinless fermions density-density interaction term by a mean-field approximation, $\hat{n}_i \hat{n}_j \simeq n_a (a^\dagger_i a_i + a^\dagger_j a_j) +(\tilde{\Delta}_{ij} a^\dagger_ia^\dagger_j + h.c) - (\tilde{\chi}_{ij} a^\dagger_i a_j + h.c)$, where $n_a = \langle a^\dagger_i a_i \rangle_a$, $\tilde{\Delta}_{ij} = \langle a_i a_j \rangle_a$, and $\tilde{\chi}_{ij} = \langle a^\dagger_j a_i \rangle_a$. This approximation may be more accurate for the regime where $J \ll t$. Notice that previously defined mean-field parameters are related to these through their complex conjugates as $\Delta_{ij} = \tilde{\Delta}_{ij} + \tilde{\Delta}^\ast_{ij}$ and $\chi_{ij} = \tilde{\chi}_{ij} + \tilde{\chi}^\ast_{ij}$. This leads to the fermion part of the Hamiltonian in real space,

\begin{eqnarray}
H_s = - \sum_{l}^r \sum^{L-1}_{j = 1} (-\omega_l e^{i\phi_l} a^\dagger_ja_{j+l} + \Delta_l e^{i \theta_l} a^\dagger_j a^\dagger_{j+l} + h.c) \\ \nonumber
+ \mu \sum^{L-1}_{j = 1} a^\dagger_j a_j
\end{eqnarray}

\noindent where we defined $ \omega_l e^{i\phi_l} = t(1 + m_l) \delta_{l1}/2 + J m_l \tilde{\chi}_l/(4 l^3 d_l^3)$, $\Omega_l e^{i \theta_l} = -t(1 - m_l) \delta_{l1}/2 + J m_l \tilde{\Delta}_l/(4 l^3 d_l^3)$, and $\mu = J n_a \sum_{l =1}^r m_l/(2 l^3 d_l^3)$.

Let us assume a closed one-dimensional lattice chain with periodic boundary conditions and make the Fourier transform $a_j = \frac{1}{\sqrt{L}} \sum_k a_k e^{ikj}$ with $k =2 \pi l/L$, where $l$ is an integer representing the lattice cordinates. Now the Hamiltonian in momentum space can be written as $H_s = \Psi^\dagger_k H_s(k)\Psi_k$ in the Nambu-spinor basis $\Psi^\dagger_k = (a^\dagger_k, a_{-k})$, where

\begin{eqnarray}
H_s(k) = \left(
        \begin{array}{cc}
          g_k + e_k & -if_k + h_k \\
          if_k + h_k & g_k-e_k \\
        \end{array}.
      \right),
\end{eqnarray}

\noindent Here we define momentum dependent interaction parameters, $g_k = \sum_l \omega_l \sin(\phi_l) \sin(kld_l)$, $e_k = \mu - \sum_l \omega_l \cos(\phi_l) \cos(kld_l)$, $f_k = \sum_l \Omega_l \cos(\theta_l) \sin(kld_l)$, and $h_k = \sum_l \Omega_l \sin(\theta_l) \sin(kld_l)$. Notice that the effective Hamiltonian for the spinless fermion sector has the form $H_s(k) = g_k \mathbb{1} + h_k \sigma_x + f_k \sigma_y + e_k \sigma_z$, where $\sigma$'s are components of usual Pauli matrices and $\mathbb{1}$ is the identity matrix. This is one of the most general Hamiltonians responsible for non-trivial topological quantum states~\cite{TPM1, TPM2, TPM3}. The Hamiltonian can be diagonalized by using usual Bogoliubov transformation through a new fermionic quasiparticles, represented by the operator $\gamma_k = e^{i\phi_1} \cos(\theta_k/2) a_k+ e^{i\phi_2} \sin(\theta_k/2) a^\dagger_{_k}$. The resulting diagonalized Hamiltonian has the form,

\begin{eqnarray}
H_s = \sum_{k, \lambda} E_{k\lambda} \gamma^\dagger_{k \lambda} \gamma_{k \lambda},
\end{eqnarray}

\noindent where the eigenvalues $E_{k\pm} = g_k \pm \sqrt{e_k^2 + f_k^2 + h_k^2}$. This allows us to calculate the previously defined local mean-field parameters, $\Delta_l = \langle a_l^\dagger a^\dagger_0 + h. c \rangle_a \equiv \tilde{\Delta}_l + \tilde{\Delta}^\ast_l$,

\begin{eqnarray}
\tilde{\Delta}_l = \sum_k F_k e^{i(kld_l + \phi_k)},
\end{eqnarray}

\noindent where,
\begin{eqnarray}
F_k =  \frac{1}{2}\biggr[\sum_\lambda n_{k\lambda}-1\biggr ] \sin(\theta_k),
\end{eqnarray}

\noindent and $ \chi_l = \langle a_l^\dagger a_0  + h. c \rangle_a \equiv \tilde{\chi}_l + \tilde{\chi}^\ast_l$,

\begin{eqnarray}
\tilde{\chi}_l = \sum_k G_k e^{ikld_l},
\end{eqnarray}

\noindent where,

\begin{eqnarray}
G_k =  \cos^2(\theta_k/2) n_{k+} - \sin^2(\theta_k/2) n_{k-} + \sin^2(\theta_k/2).
\end{eqnarray}

\noindent Here we defined, $\phi_k \equiv \phi_1 -\phi_2 = \arctan[f_k/h_k]$ and $n_{k\lambda} = (e^{ E_{k\lambda}/k_BT} + 1)^{-1}$. The parameter $\theta_k$ is determined by $\tan (\theta_k) = \sqrt{2} [f_k \sin(2 \phi_1) + h_k \cos(2 \phi_1)]/e_k$ with the constraint $\phi_1 + \phi_2 = \pi/4$.

\section{VII. Self-Consistent Equations}

The spinless particle correlations $\Delta_l$ and $\chi_l$ must be calculated through the self-consistent equations (12), (17), and (19). Using these three equations, first we write down system parameters, $e_k$, $f_k$, $g_k$, $h_k$, and $J^e(k)$ in terms of the functions $G_k$ and $F_k$, and $m_k$.

\begin{eqnarray}
e_k = \frac{J}{8 d^3_l} \sum_{k^\prime,k^{\prime \prime}} G_{k{^\prime}} m_{k^{\prime \prime}} \biggr\{ 8 \Re\Gamma_{k^{\prime \prime}} -  \Re \Gamma_{q+k} - \Re \Gamma_{q-k} \biggr \} \nonumber \\
- \frac{t}{2} \biggr[1 + \sum_{k^\prime} m_{k^\prime} \cos(k^\prime d_l) \biggr] \cos(kd_l),
\end{eqnarray}

\begin{eqnarray}
f_k = \frac{J}{8 d^3_l} \sum_{k^\prime,k^{\prime \prime}}  F_{k{^\prime}} \sin (\phi_{k^\prime}) m_{k^{\prime \prime}} [\Re \Gamma_{q+k} + \Re \Gamma_{q-k}] \nonumber \\
+ \frac{J}{8 d^3_l} \sum_{k^\prime,k^{\prime \prime}}  F_{k{^\prime}} \cos (\phi_{k^\prime}) m_{k^{\prime \prime}} [\Im \Gamma_{q+k} + \Im \Gamma_{q-k}] \nonumber \\
- \frac{t}{2} \biggr[1 - \sum_{k^\prime} m_{k^\prime} \cos(k^\prime d_l) \biggr] \sin(kd_l),
\end{eqnarray}

\begin{eqnarray}
g_k = \frac{J}{8 d^3_l} \sum_{k^\prime,k^{\prime \prime}}  G_{k{^\prime}} m_{k^{\prime \prime}} [\Re \Gamma_{q+k} - \Re \Gamma_{q-k}] \nonumber \\
+ \frac{t}{2} \biggr[1 + \sum_{k^\prime} m_{k^\prime} \sin(k^\prime d_l) \biggr] \sin(kd_l),
\end{eqnarray}

\begin{eqnarray}
h_k = \frac{J}{8 d^3_l} \sum_{k^\prime,k^{\prime \prime}}  F_{k{^\prime}} \cos(\phi_{k^\prime}) m_{k^{\prime \prime}} [\Re \Gamma_{q+k} - \Re \Gamma_{q-k}] \nonumber \\
+ \frac{J}{8 d^3_l} \sum_{k^\prime,k^{\prime \prime}}  F_{k{^\prime}} \sin(\phi_{k^\prime}) m_{k^{\prime \prime}} [\Im \Gamma_{q-k} - \Im \Gamma_{q+k}] \nonumber \\
- \frac{t}{2} \biggr[1 - \sum_{k^\prime} m_{k^\prime} \sin(k^\prime d_l) \biggr] \sin(kd_l),
\end{eqnarray}

\noindent and

\begin{eqnarray}
J^e(k) = 2 t \sum_{k^\prime} \biggr[2 G_{k^\prime} \cos(k^\prime d_l) + F_{k^\prime} \cos(k^\prime d_l +\phi_{k^\prime})\biggr]
\nonumber \\
\times \cos(kd_l) + \frac{J}{d^3_l} \biggr(\sum_{k^\prime} G_{k{^\prime}}\biggr)^2 \Re \Gamma_k,
\end{eqnarray}

\noindent where we have defined $q = k^\prime + k^{\prime \prime}$, and $\Re \Gamma_k$ and $\Im \Gamma_k$ are real and imaginary parts of the third order polylogarithm $\Gamma_k \equiv Li_3[e^{ikd_l}] = \sum_{l} e^{ikld_l}/l^3$. Equations (20-24), together with Eqs. (12), (17), and (19), allows one to self-consistently solve for the momentum dependent functions $G_k$, $F_k$, and $m_k$.

\section{VIII. Correlation Functions}

Within the slave-spin approach, the density correlation function $\langle \hat{n}_{ic} \hat{n}_{jc} \rangle$, where $\hat{n}_{ic} = \sum_s \hat{n}_{i,s}$ is the total particle number operator at site $i$ with $\hat{n}_{i,s} = c^\dagger_{i,s}c_{i,s}$, the spin correlation function $\langle \vec{S}_i \cdot \vec{S}_j \rangle$, and the pairing correlation function $\langle \vec{P}_i \cdot \vec{P}_j \rangle$, all can be written in terms of spinless fermion density correlation $D_l$, pseudo spin correlation $m_l$, and the average particle density parameter $n = \langle \hat{n}_{ic} \rangle$. These correlation functions have the forms, $\langle \hat{n}_{ic} \hat{n}_{jc} \rangle = (1-2 n_a + D_l) m^z_l$, $\langle \vec{S}_i \cdot \vec{S}_j \rangle = D_l m_l/4$, and $\langle \vec{P}_i \cdot \vec{P}_j \rangle = \langle \vec{S}_i \cdot \vec{S}_j \rangle - n_am_l/2+m_l/4$, where the on-site spinless fermion occupation number $n_a = \langle a^\dagger_i a_i \rangle$. Here $D_l$ is the spinless fermion density correlation function defined through $D_l = \langle \hat{n}_i \hat{n}_j \rangle$. Using bi-linear decoupling for the four spinless fermion operators, this spinless density correlations can be written as $D_l = n^2_a +  |\tilde{\Delta}_l|^2 - |\tilde{\chi}_l|^2$. Finding these correlation functions are still a challenging task due to the self-consistency and the momentum dependence of the equations.

In order to get an understanding of the general solution, first we consider a special case where the spiral wave-vector for pseudo spin sector $\vec{Q} = 0$ at zero temperature. For this case, the mean-field parameter $m_k$ is momentum independent, thus the structure of the mean-field parameters $\Delta_l$ and $\chi_l$ are determined by the spinless fermion sector represented by the Kitaev type Hamiltonian in Eq. (14) alone. The mean-field parameter $m_k$ simply renormalizes the functions $G_k$ and $F_k$.  The correlation effects of both fermionic and bosonic Kitaev type long range models have been investigated before~\cite{CP1,CP2, CP3, CP4, CP5}. The stability of topological phases upon changing the exponent of the long-range interaction has been studied in ref.~\cite{CP4}. Quantum correlation of pure Ising and XY spin models and entanglement properties as a function of the long-range interaction have been studied in ref.~\cite{CP2} and ref.~\cite{CP3}, respectively. Studies on correlation functions of Kitaev type models in ref.~\cite{CP1} and ref.~\cite{CP5} find hybrid exponential and algebraic behavior. Therefore, the momentum dependent functions $G_k$ and $F_k$ in spinless fermion correlations  at this specific $\vec{Q} = 0$ case can be written in the form~\cite{CP1},

\begin{eqnarray}
G_k = \sum_l \biggr(\frac{A_g (-1)^l e^{-\xi_g l}}{\sqrt{l}} + \frac{B_g}{l^4} \biggr) \cos(kld_l) \nonumber \\
F_k \cos(\phi_k) = \sum_l \biggr(\frac{A_f (-1)^l e^{-\xi_f l}}{\sqrt{l}} + \frac{B_f}{l^3} \biggr) \cos(kld_l) \nonumber \\
F_k \sin(\phi_k) = -\sum_l \biggr(\frac{A_f (-1)^l e^{-\xi_f l}}{\sqrt{l}} + \frac{B_f}{l^3} \biggr) \sin(kld_l)
\end{eqnarray}

\noindent where $A_{g,f}$, $B_{g,f}$, and $\xi_{g,f}$ are interaction dependent constants. Assuming these solutions, we solve our self consistent equations variationally for six variational parameters $A_{g,f}$, $B_{g,f}$, and $\xi_{g,f}$. First, we insert these variational $G_k$ and $F_k$ in our self consistent equations, and then derive new expression for these functions using Eqs. (17) and (19). Second, by expanding the variational functions and newly constructed functions as powers of momenta $k$ and equating the first three non-zero orders for both functions, we construct six equations for variational parameters $A_{g,f}$, $B_{g,f}$, and $\xi_{g,f}$. As we are equating the non-zero lowest order three coefficients of power series, our result may valid only for the low energy sector. We numerically solve the six variational equations and find that the solutions exist only in the $1/\xi_{g,f} \rightarrow 0$ limit. This indicates the absence of exponential behavior in correlation functions. This result is not surprising as exponential behaviour originates from the massless edge modes and algebraic tail originates from the bulk of the system~\cite{CP1}. As we have used periodic boundary conditions, our system does not produce edge modes. Without, periodic boundary conditions, the correlation functions show both power law and exponential behavior. The algebraic behaviour of the correlation functions for our closed spinless fermions sector is consistent with the findings of~\cite{CP1, CP5}. However, the effective interaction parameters in our model needs to be calculated self consistently. As a result, we have a set of self-consistent equations to be solved with combination of pseudo spin part of the Hamiltonian. Thus, the behavior of correlation functions of the $t-J$ model has different parameter dependence. Further, we justify that the absence of exponential behaviour by using purely exponential and algebraic decay variational functions and solving the variational equations for  $A_{g,f}$ and $\xi_{g,f}$.

Now we have established the form of the solutions for $G_k$ and $F_k$ for a special case of pseudo spin wave-vector $\vec{Q} = 0$. Before we relax the constraint $\vec{Q} = 0$, we argue that the general form of the solutions do not change even for non-zero values of $\vec{Q}$. The quantitative effect of non-zero $\vec{Q}$ can be included by a proper choice of a variational function for the pseudo spin correlation function $m_l$. The physically relevant pseudo spin wave-vector $\vec{Q}$ is not known, however $\vec{Q}$ can be fixed numerically in our calculation scheme by using the conformal field theoretical result for the spin correlation function~\cite{CFT}. This allows us to write the z-component of the pseudo-spin correlation function $m^z_l = (12/D_l)[1/(\pi l d_l)^2 + B \cos(2 k_F l d_l) \sqrt{\ln(l d_l)}/(l d_l)^{1+K}]$, where $k_F$ is the Fermi momentum. In addition to the variational parameters $B_g$ and $B_f$, this add two additional momentum independent variational parameters $B$ and $K$ to our self-consistent calculations. Within our numerical calculation scheme, we insert these variational ansatz for $G_k$, $F_k$, and $m^z_l$ into our self consistent equations and solve for the variational parameters $B_g$, $B_f$, $B$, and $K$ variationally.

 \begin{figure}
\includegraphics[width=\columnwidth]{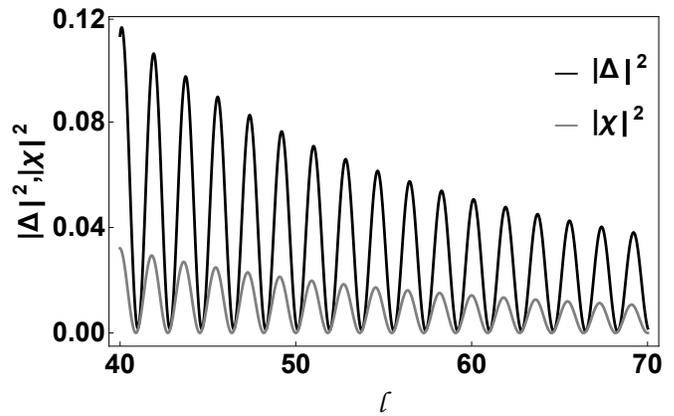}
\caption{(color online) Absolute values of spinless fermion pairing correlation and density correlation, $|\Delta|^2$ and $|\chi|^2$ at particle density $n = 0.2$ and interaction $J = 0.25 t$.}\label{SFCF}
\end{figure}

\begin{figure}
\includegraphics[width=\columnwidth]{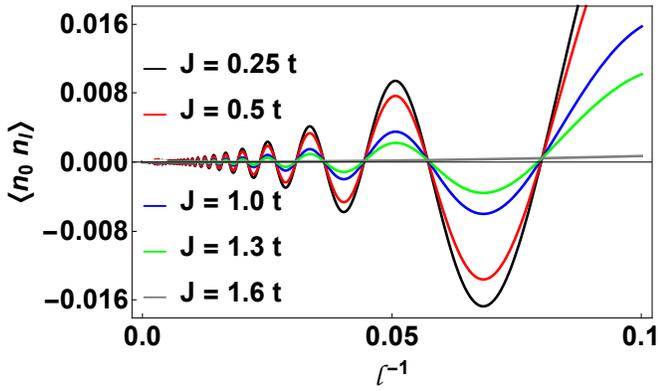}
\caption{(color online) The particle density correlation as a function of inverse lattice coordinate $l^{-1}$ for a series of interaction parameters $J$. The particle density is fixed to be $n = 0.2$.}\label{DCa}
\end{figure}

\begin{figure}
\includegraphics[width=\columnwidth]{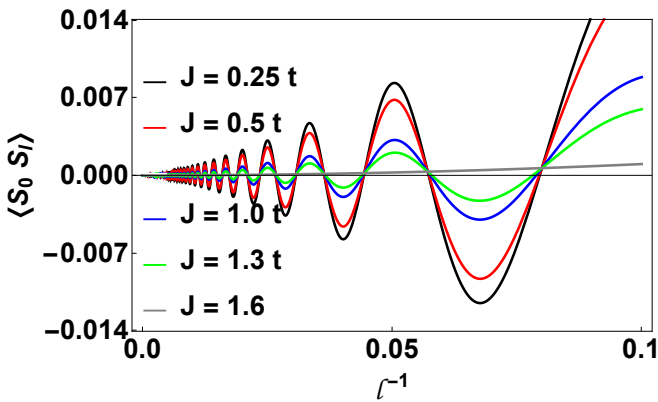}
\caption{(color online) The spin correlation as a function of inverse lattice coordinate $l^{-1}$ for a series of interaction parameters $J$. The particle density is fixed to be at $n = 0.2$.}\label{SCa}
\end{figure}

\begin{figure}
\includegraphics[width=\columnwidth]{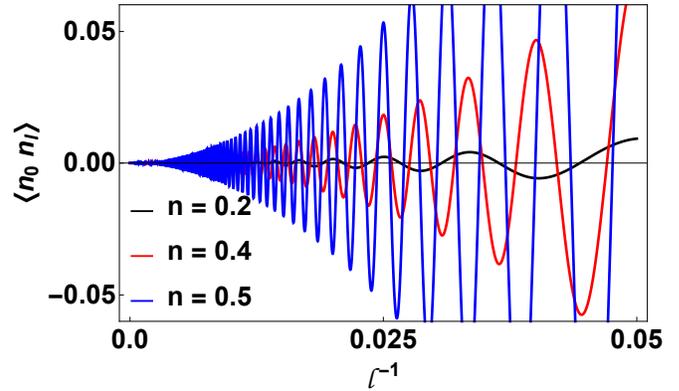}
\caption{(color online) The particle density correlation as a function of inverse lattice coordinate $l^{-1}$ for a series of density parameters $n$. The interaction parameter is fixed to be $J = 0.25 t$.}\label{DCb}
\end{figure}

\begin{figure}
\includegraphics[width=\columnwidth]{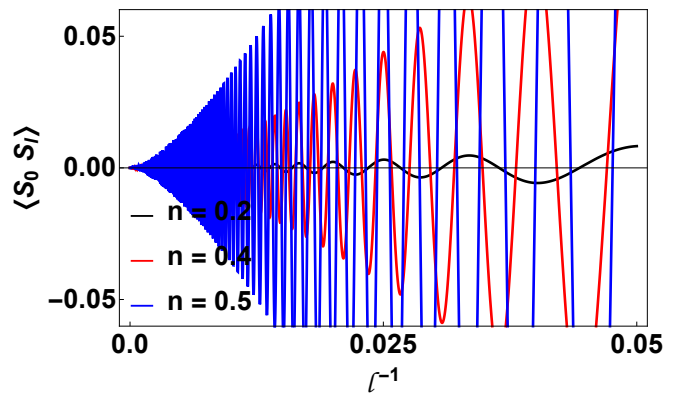}
\caption{(color online) The spin correlation as a function of inverse lattice coordinate $l^{-1}$ for a series of density parameters $n$. The interaction parameter is fixed to be at $J = 0.25 t$.}\label{SCb}
\end{figure}

Our results are summarized in FIGS.~\ref{SFCF}, \ref{DCa}, \ref{SCa}, \ref{DCb} and \ref{SCb}. Notice that the \emph{integer} lattice site separation $l$ enters in our formalism through the parameter $ld_l$. In our calculations, we set the range of the dipole-dipole interaction $r \rightarrow ld_l$ and set maximum number of lattice sites to be $l_m = 70$. We treated the lattice constant $d_l$ as a length scale and set it to be one. Our results are presented as functions of lattice separation $l$ or inverse lattice separation $l^{-1}$. The non-integer $l$ values are only guide to the eye and formally do not contain any physical information. The term $(-1)^l$ for non-integer values of $l$ in Eq. (25) has no effect on our final results as this term vanishes due to the fact $1/\xi_{g,f} \rightarrow 0$. Notice that the results are valid only for long-distance limit due the ansatz we used for $m_l$ to fix $\vec{Q}$, which is valid only for the asymptotic limit.

As a demonstration for the structure of the spinless particle correlation functions, we plot $|\Delta|^2$ and $|\chi|^2$ as a function of lattice coordinate for two representative values of the average particle density $n = 0.2$ and interaction parameter $J = 0.25 t$ in FIG.~\ref{SFCF}. As can be seen from the figure, both of these correlation functions show oscillatory behavior with the same period fixed by the average particle density. The FIG.~\ref{DCa} and FIG.~\ref{SCa} show the physical particle density correlation $\langle \hat{n}_{ic} \hat{n}_{jc} \rangle \equiv \langle n_{0} n_{l} \rangle$ and spin correlation function $\langle \vec{S}_i \cdot \vec{S}_j \rangle \equiv \langle S_0 S_l \rangle$ as a function of inverse lattice coordinates $l^{-1}$ for a series of interaction parameters $J/t$. For the demonstration purpose, here we set $i \rightarrow 0$ and $j \rightarrow i+l$. Notice that both correlation functions as function of inverse lattice coordinate $l^{-1}$ shares the same qualitative features for all $J/t$. They both shows the same period, but different amplitudes. As evidence from the earlier predictions~\cite{CFT}, the oscillatory period remains constant due to the fact that density $n \propto k_F$ is fixed for these figures. Notice that the oscillatory behaviour seen here for $J < 1.6 t$ has already been predicted for the nearest neighbor only $t-J$ model also. Using Monte Carlo simulations~\cite{MCS}, variational approaches~\cite{VAP}, and finite-size scaling~\cite{FSS}, it has been shown that there are no qualitative changes in the static properties of the regular nearest neighbor only $t-J$ model for $J \leq 2 t$. We find that the correlation functions start to demonstrate qualitatively different behavior for larger values of $J  > 1.3 t$. For larger values of $J > 1.3 t$, both density and spin correlation functions starts to show power low decay as shown in FIG.~\ref{DCa} and FIG.~\ref{SCa}. This different behavior may be attributed to two reasons. First, it may be due to the phase separation of particle and holes as suggested by the studies of nearest neighbor only $t-J$ model~\cite{MCS,VAP, VAPM, rfn3}, although in these studies the phase separation transition has been found to occur closer to $J = 3 t$~\cite{rfn1, rfn2}. Second, as the slave-boson theories are accurate for stronger coupling limits ($U \gg t$) of the Hubbard model, our calculation may not be valid for $ J \gg t$ limit of our $t-J$ model. As the $J \ll t$ limit is equivalent to the strong coupling limit of the Hubbard model, our results may not be accurate for larger $J$ values.

The effect of density variation on the correlation functions are shown in FIG.~\ref{DCb} and FIG.~\ref{SCb}. These two figures show the particle density and spin correlation functions as a function of inverse lattice coordinates $l^{-1}$ for a series of density parameters $n$ at a fixed interaction strength $J = 0.25 t$. Notice that the different oscillatory period due to the density variation and rapid increase in amplitude as one increases the density. For other values of $J < 1.6 t$, these correlation functions share the same qualitative behavior. Due to the fact that the pairing correlation function $\langle \vec{P}_i \cdot \vec{P}_j \rangle$ is directly related to the spin correlation function and the pseudo spin correlation function $m_l$, the pairing correlations also shows qualitatively similar features as a function of lattice coordinate $l$.

The oscillatory behavior of the correlation functions can be justified by the analytical results at limiting cases~\cite{CFT, rfn1}. The spin correlation function $(1- \cos 2k_Fld_l)/8$ for a limiting case where spin-flip term and long-range part absence and at $J = 4 t$~\cite{rfn1}. The limit $J \rightarrow 0$ is the stronger coupling limit where $U \rightarrow \infty$ where charge dynamics can be considered as non-interacting spinless fermions. The charge correlation has an analytic form at this limit $\langle \hat{n}_{ic} \hat{n}_{jc} \rangle \propto [\cos (\pi l)-1]/(2 \pi^2 l^2)$~\cite{rfn1}.

In a recent density matrix renormalization (DMRG) study on one-dimensional $t-J$ model without the long-range part, different phases have been characterized by density and spin correlation functions~\cite{rfn3}. Here we find that the long-range interactions have significant influence on the ground state properties. Indeed, DMRG studies on long-range dipolar interactions, a novel metallic phase with a spin-gap has been discovered~\cite{rfn4}. Extending the DMRG scheme to study correlation functions for long-range interactions, it has been shown that transverse spin correlation function and the density correlation function has algebraic decays with exponent 8.7~\cite{rfn2}. This algebraic decay is consistent with our results at larger values of interaction parameter $J$.

\section{IX. Discussion and Conclusions}

The model studied in this paper relevant for dipolar fermions is a generalized version the original $t-J$ model defined on a lattice whose sites can be either occupied by one particle or not. This is the strongly correlated limit of the single band Hubbard model away from half filling. The spin exchange, particle repulsion, and hopping of atoms between lattice sites are all taken into account through the model. Even in the one-dimensional case of our interests, the $t-J$ model is not exactly solvable, except for extreme limiting cases with only nearest neighbor interactions. Therefore, our approximate theoretical solutions of the long-range $t-J$ model relevant for dipolar fermions serve as a valuable tool for studying the strongly correlated atoms. In addition to calculating the correlation functions, our method would allow one to investigate thermodynamic properties of the dipolar fermions.

On the experimental side, the dipolar fermionic molecules and atoms, such as $^{40}$K$^{87}$Rb, $^{27}$Na$^6$Li, $^{167}$Er, $^{53}$Cr and $^{161}$Dy, can be promising candidates for realizing a truly long-range $t-J$ model~\cite{mwf4,c1,c2,c4,c5,c6,c7}. A degenerate Fermi gas of ultracold polar molecules of potassium-rubidium, $^{40}$K$^{87}$Rb has already been brought to Fermi degeneracy~\cite{c7A}. The long range interactions further can be controlled via time-dependent dipole orientation or state-dressing~\cite{cc1,cc2}. The correlation functions discussed in this paper can be measured with currently available experimental techniques in cold gas experiments. For example, the spin and particle correlations can be detected by employing coherent microwave spectroscopy~\cite{embp2}, using spin blockade effects~\cite{edm1}, coupling atoms to light~\cite{edm2}, using quantum noise analysis techniques~\cite{edm3}, measuring the fraction of atoms residing in a lattice site due to the loss dynamics~\cite{edm4}, using cold atom microscopy~\cite{edm5}, and  applying other spectroscopic techniques~\cite{edm6, edm7,edm8,edm9,edm10,edm11} or periodic force techniques~\cite{edm12}.

In conclusion, we have developed an approximate theoretical scheme to calculate the density, spin, and pairing correlation functions for long range one-dimensional fermions subjected to a periodic optical potential. By combining a constraint free slave-spin approach with a mean field theory, we derived a set of self-consistent equations for the correlation functions. The calculated density, spin, and pairing correlation functions are related through the spinless fermion density correlation and pseudo-spin correlation originated from the slave-spin sector.

\section{X. Acknowledgments}

The author acknowledges the support of Augusta University and the hospitality of KITP at UC-Santa Barbara. A part of this research was completed at KITP and was supported in part by the National Science Foundation under Grant No. NSF PHY11-25915.

\end{document}